# Observation of the Density Minimum in Deeply Supercooled Confined Water

Dazhi Liu*, Yang Zhang*, Chia-Cheng Chen‡, Chung-Yuan Mou‡, Peter H Poole§,

Sow-Hsin Chen*†

* Department of Nuclear Science and Engineering, Massachusetts Institute of Technology, Cambridge, MA, 02139; ‡ Department of Chemistry, National Taiwan University, Taipei, Taiwan 106; § Department of Physics, St. Francis Xavier University, Antigonish, Nova Scotia, B2G 2W5, Canada

†To whom correspondence should be addressed at: Department of Nuclear Science and Engineering, Massachusetts Institute of Technology, Cambridge, MA, 02139. E-mail: sowhsin@mit.edu.

**ABSTRACT**

**Small angle neutron scattering (SANS) is used to measure the density of heavy water contained in 1-D cylindrical pores of mesoporous silica material MCM-41-S-15, with pores of diameter of 15±1 Å. In these pores the homogenous**

**nucleation process of bulk water at 235 K does not occur and the liquid can be supercooled down to at least 160 K. The analysis of SANS data allows us to determine the absolute value of the density of $D_2O$ as a function of temperature. We observe a density minimum at 210±5 K with a value of 1.041±0.003 g/cm$^3$. We show that the results are consistent with the predictions of molecular dynamics simulations of supercooled bulk water. This is the first experimental report of the existence of the density minimum in supercooled water.**

TEXT

Of the many remarkable physical properties of liquid water [1], the density maximum is probably the most well known. The density maximum of $H_2O$ at $T_{max}$=277 K (284 K in $D_2O$) is one of only a few liquid-state density maxima known [2], and the only one found to occur in the stable liquid phase above the melting temperature. Water's density maximum is a dramatic expression of the central role played by hydrogen bonding in determining the properties of this liquid: as temperature T decreases through the region of the density maximum, an increasingly organized and open four-coordinated network of hydrogen bonds expands the volume occupied by the liquid, overwhelming the normal tendency of the liquid to contract as it is cooled.

The density of bulk supercooled liquid water decreases rapidly with T before the onset of homogeneous crystal nucleation (at approximately 235 K) precludes further measurements. The density curve of ice Ih lies below that of the liquid, and almost certainly sets a lower bound on the density that the supercooled liquid could attain if nucleation were avoided, since ice Ih represents the limiting case of a perfectly ordered tetrahedral network of hydrogen bonds. Significantly, the expansivity of ice Ih in this T range is positive [3], i.e. the density increases as T decreases (see Fig. 2). The low density amorphous (LDA) ice that forms from deeply supercooled liquid water at the (in this case extremely weak) glass transition, approaches very closely the structure of a "random tetrahedral network" (RTN), and exhibits a number of ice-like properties, including a "normal" (i.e. positive) expansivity [4]. If the structure of deeply supercooled water also approaches that of a RTN, it is therefore possible that a density *minimum* occurs in the supercooled liquid [5].

Consistent with this possibility, a number of recent molecular dynamics (MD) computer simulation studies predict that a density minimum occurs in water ($H_2O$) [5-8]. These studies achieve deep supercooling without crystal nucleation due to the small system size and short observation time explored, compared to experiment. In the literature the five-site transferable interaction potential (TIP5P) for water is considered to be the most accurate model for reproducing experimental data when used with a simple spherical cutoff for the long-ranged electrostatic interactions in MD simulation [9]. The ST2 potential is widely used in simulation of water since early 1970s [10]. As shown in Fig. 2, the TIP5P-E [7] model of water exhibits a density minimum at a temperature $T_{min}$ approximately 70 K below $T_{max}$ at atmospheric pressure. The ST2 model also predicts that a density minimum occurs at approximately this temperature [6,8].

Density minima in liquids are even more rare than density maxima. We are aware of reports of density minima in only a few liquid systems, such as Ge-Se mixtures [11]. Confirming the existence of a density minimum in water would reveal much about the supercooled state of this important liquid. Its occurrence would signal the reversal of the anomalies that set in near the density maximum; i.e. that mildly supercooled water is anomalous, but that deeply supercooled water "goes normal". Observing a density minimum would also have significant implications for the possibility that a liquid-liquid phase transition (LLPT) occurs in supercooled water [1,12], along the same lines as was recently argued for vitreous silica [13].

In this report, we present the first experimental observation of a density minimum in supercooled water (D$_2$O), confined in the nanochannels of mesoporous silica, occurring at T$_{min}$=210±5 K with a density value of 1.041±0.003 g/cm$^3$ as shown in Fig. 1. Our sample is a fully hydrated (D$_2$O) MCM-41-S-15 powder, which is made of a micellar templated mesoporous silica matrix and has 1-D cylindrical pores arranged in a 2-D hexagonal lattice [14]. In this experiment, we choose the material with a pore diameter of 15±1 Å because the differential scanning calorimetry data show no freezing peak down to 160 K for the fully hydrated sample. The SANS diffraction pattern from the sample consists of two parts: inter-grain interfacial scattering and a Bragg peak coming from the 2-D hexagonal internal structure of the grains. Based on the detailed analysis (presented in Method section), we find that the height of Bragg peak is related to the scattering length density (sld) of D$_2$O inside the silica pores and the sld of D$_2$O is proportional to its mass density $\rho_{D_2O}^m$. Hence we are able to determine the density of water (D$_2$O) by measuring the temperature dependent neutron scattering intensity I(Q). The highest value corresponds to the known density maximum of D$_2$O at about 284 K. Most significantly, we find a density minimum situated at T$_{min}$=210±5 K with a value of 1.041±0.003 g/cm$^3$.

**Discussion**

Fig. 2 compares the density minimum found here with that found in Paschek's MD simulation study of TIP5P-E [7]. The correspondence in T with our experimental data is excellent: for TIP5P-E, $T_{max}-T_{min}$ is approximately 70 K, compared with 80 K in the present study. Note also that the ratio of the maximum to minimum density is 1.05 for TIP5P-E, compared with 1.06 for our data.

The minimum density we find (1.041 g/cm$^3$) also compares well with the density of LDA ice, to which the deeply supercooled liquid will transform at the glass transition, if crystallization is avoided. The density of H$_2$O LDA ice is 0.94 g/cm$^3$ [4], corresponding to approximately 1.04 g/cm$^3$ for D$_2$O LDA ice, assuming a 10.6% density difference [15]. Also of note is that a super-Arrhenius to Arrhenius dynamic crossover phenomenon has been experimentally observed in this confined water (H$_2$O) system at 225 K [14]; hence $T_{min}$ occurs in a regime of strong liquid behavior, below this crossover. Together, these observations strongly suggest that the structure of water below $T_{min}$ is approaching that of a fully-connected, defect-free hydrogen bond network, in which the anomalies of water, so prominent near the melting temperature, disappear.

The finding of a density minimum also has significant implications for the proposal that a liquid-liquid phase transition (LLPT) occurs in supercooled water [1,12]. These implications arise because of formal relationships that exist between density anomalies and response functions, such as the isothermal compressibility $K_T$ and the isobaric specific heat $C_P$ [16]. For example, $K_T$ is known experimentally to *increase* with decreasing T near $T_{max}$, and it has been shown that this must be true in any system upon crossing a line of density maxima having negative slope in the T-P plane. When applied to the vicinity of a density minimum, these same thermodynamic relations predict that $K_T$ must be *decreasing* with T at $T_{min}$, under the physically plausible assumption that the line of density minima also has a negative slope in the T-P plane. In combination, these two constraints on the behavior of $K_T$ mean that $K_T$ must have a maximum between $T_{min}$ and $T_{max}$. A $K_T$ maximum implies that other response functions, such as the specific heat $C_P$, will also attain extrema in the range between $T_{min}$ and $T_{max}$. A $C_P$ peak in this range is consistent with the dynamical crossover observed at 225 K [14]. Furthermore, the occurrence of an inflection point in our data for the T dependence of the density (see Fig. 1) directly establishes that the thermal expansion coefficient $\alpha_p = -1/\rho (\partial \rho / \partial T)_P$ has a minimum between $T_{min}$ and $T_{max}$. We know for fact that $\alpha_p$ has a peak when we cross the Widom line above the liquid-vapor transition of steam [17]. This reinforces the plausibility that there is a Widom line emanating from the liquid-liquid critical point in supercooled water passing between $T_{min}$ and $T_{max}$, as was indicated in our previous experiment which detects the dynamic crossover phenomenon at 225 K at ambient pressure [18].

This pattern of thermodynamic behavior, in which density anomalies bracket extrema in response functions at ambient pressure, is entirely consistent with that found in MD simulations of TIP5P-E and ST2, both of which exhibit a LLPT at elevated pressure (see in particular Fig. 2 of Ref. [8]). Conversely, MD models of water in which a density minimum has not been observed are also models in which a LLPT has not been identified. In the T-P plane, the lines of response function extrema necessarily meet at the critical point of the LLPT, if it exists. The present results, combined with those of Ref. [14], demonstrate that the temperature window in which these extrema occur can be identified and accessed at ambient pressure, and that the system remains an ergodic liquid throughout this range. Therefore, if the critical point of a LLPT exists at higher pressure, there is considerable promise that it too occurs in the liquid regime, and can be detected by mapping the equation of state for the density via SANS experiments of the kind described here, conducted at higher pressure.

Overall, our results demonstrate that SANS is a powerful method for determining the average density of $D_2O$ in cylindrical pores of MCM-41-S-15 silica matrix. It remains an open question whether the density minimum we find in confined water can be confirmed in bulk water. However, given the importance of confined water, particularly in biological systems, our demonstration of the disappearance of water's anomalies in confinement below 210 K has broad implications for understanding the low temperature properties of a wide range of aqueous microstructured systems, as well as bulk water itself.

**Methods**

**Neutron experiment**

SANS experiments were performed at NG3, a 40-m SANS spectrometer, in the NIST Center for Neutron Research (NCNR). The incident monochromatic neutrons have an average wave length of $\lambda=5$ Å with a fractional spread of $\Delta\lambda/\lambda=10\%$. The sample to detector distance is fixed at 6m, covering the range of magnitudes of neutron wave vector transfer (Q) from 0.008 Å$^{-1}$ to 0.40 Å$^{-1}$. This Q range covers the high Q part of the interfacial scattering between different grains and the Bragg peak due to the hexagonal array of silica pores within a grain. The amplitude of the latter is used as an indicator of the density of water in the sample.

**Sample preparation**

Our sample consists of a fully hydrated ($D_2O$) MCM-41-S-15 powder, which is made of micellar templated mesoporous silica matrices and has 1-D cylindrical pores arranged in a 2-D hexagonal lattice [14]. We synthesize MCM-41-S-15 by reacting pre-formed $\beta$-zeolite seeds (formed with tetraethylammonium hydroxide(TEAOH)) with decyltrimethylammonium bromide solution ($C_{10}$TAB, Acros), then transferring the mixture into an autoclave at 150℃ for 18 hours. Solid samples are then collected by filtration, washed with water, dried at 60℃ in air overnight, and calcined at 560℃ for 6 hours. The molar ratios of the reactants are $SiO_2$ : NaOH : TEAOH : $C_{10}$TMAB : $H_2O$ ＝1 : 0.075 : 0.285 : 0.22 : 52.

The sample is hydrated by exposing it to water vapor in a closed chamber until it reaches the full hydration level of 0.5 gram $D_2O$/1 gram silica. We make measurements at 14 temperatures between 160 K and 290 K in a step of 10 K to monitor the variation of the density in the supercooled region (see Fig. 3).

**Data analysis**

The powder sample of MCM-41-S-15 we use in the experiment consists of crystallites, or grains (approximately spherical) of the order of micron size. Each grain is made up of a 2-D hexagonal matrix of parallel cylindrical silica pores with an inter-pore distance $a$. After full hydration, all the pores are filled with water ($D_2O$), which has a considerably different scattering length density (a factor two larger) from that of the silica matrix. The direction of the cylindrical axis in each grain is randomly distributed in space. The diffraction pattern from the sample therefore consists of two parts: (i) inter-grain interfacial scattering, the Q-dependence of which follows a power law; and (ii) a Bragg peak at $Q_1=2\pi/a$ coming from the 2-D hexagonal internal structure of the grains.

Fig. 3 shows a peak situated at $Q_1=0.287$ Å$^{-1}$, which corresponds to the center-to-center distance between the water columns $a=2\pi/Q_1 \approx 21.9$ Å. In the range of 0.1-0.2 Å$^{-1}$, a straight line in log-log scale represents the asymptotic part of the interfacial scattering.

In a SANS experiment, the measured Q-vector is essentially perpendicular to the incident neutron direction. For a scattering unit (particle), which is a long cylinder with a small circular cross section (such as the present case), the scattering geometry essentially selects out only those cylinders which happen to lie with their cylindrical axes parallel to the incident neutron direction. Consequently the direction of the measured Q-vector is nearly perpendicular to the cylindrical axis. With this understanding, the neutron scattering intensity distribution I(Q) is given by $I(Q) = nV_p^2(\Delta\rho)^2\overline{P}(Q)S(Q)$, where n is the number of scattering units (water cylinders) per unit volume in the sample, $V_p$ the volume of the scattering unit, $\Delta\rho = \rho_p - \rho_e$ the difference of scattering length density (sld) between the scattering unit $\rho_p$ and the environment $\rho_e$, $\overline{P}(Q)$ the normalized particle structure factor (or form factor) of the water cylinder, and $S(Q)$ the inter-particle structure factor (of a 2-D hexagonal lattice). Note that the sld of the confining material MCM-41-S-15 is $3.618 \times 10^{10}$ cm$^{-2}$ and is approximately independent of temperature in the temperature range we study (as is evidenced by the fact that the position of $Q_1$ changes by less than 0.5% for the entire temperature range studied, see Table 1). The sld of the of scattering unit $\rho_p$ can be rewritten as $\rho_p = \rho_{D_2O}^m N_A \sum_i b_i / M_w$, where $N_A$ is Avogadro's number, $M_w$ the molecular weight of D$_2$O, $b_i$ the coherent scattering length of the $i$-th D$_2$O molecule in the scattering unit, and $\rho_{D_2O}^m$ the mass density of D$_2$O. The sld of the environment $\rho_e$ has been determined by a separate contrast variation experiment. Based on the above relations, we find that all the variables in the expression for I(Q) are independent of temperature except for $\Delta\rho$ since it involves a temperature dependent parameter $\rho_{D_2O}^m$. Hence we are able to determine the density of water (D$_2$O) by measuring the temperature dependent neutron scattering intensity I(Q).

The normalized particle structure factor $\overline{P}(Q)$ of a long (QL>2π) cylinder is given by $\overline{P}(Q) = \pi/QL(2J_1(QR)/QR)^2$ [19], where *L* and *R* represent the length and the radius of the cylinder respectively, and $J_1(x)$ is the first-order Bessel function of the first kind. As an example, the form of $\overline{P}(Q)$ is depicted as a green solid line in Fig. 4A. The structure factor S(Q) of a perfect 2-D hexagonal lattice is a series of delta functions (Bragg peaks) situated at $Q_1=2\pi/a$, $Q_2=2\sqrt{3}\pi/a$, … where *a* is the length of the primitive lattice vector of the 2-D hexagonal lattice. All the Bragg peaks will be broadened due to defects of the lattice and the finite size of the grains. The broadening can be well approximated by a Lorentzian function. The black solid line in Fig. 4A shows the S(Q) in our model. Therefore, the neutron intensity we measured in the $Q_1$ peak region (0.2-0.4 Å$^{-1}$), after subtraction of the interfacial scattering, is expressed as,

$$I(Q) = nV_p^2 \left( \frac{N_A \sum_i b_i}{M_w} \rho_p^m - \rho_e \right)^2 \frac{\pi}{QL} \left( \frac{2J_1(QR)}{QR} \right)^2 \left( C \frac{\frac{1}{2}\Gamma}{(Q-\frac{2\pi}{a})^2 + (\frac{1}{2}\Gamma)^2} \right), \quad (1)$$

where Γ is the FWHM and C a temperature-independent constant. Combining all constants, we obtain

$$I(Q) = C_1 \frac{J_1(QR)^2}{Q^3 R^2} \left( \frac{\frac{1}{2}\Gamma}{(Q - \frac{2\pi}{a})^2 + (\frac{1}{2}\Gamma)^2} \right) , \qquad (2)$$

where the new prefactor $C_1 = 4CnV_p^2 \frac{\pi}{L} \left( \rho_p^m - \left( M_w / N_A \sum_i b_i \right) \rho_e \right)^2 \propto (\rho_{D_2O}^m - C_0)^2$, and $C_0 = \left( M_w / N_A \sum_i b_i \right) \rho_e =$ 0.6273 g/cm$^3$ determined by $\rho_e$ of MCM-41-S-15.

By fitting the model described above to our data for neutron intensity I(Q), the parameters $C_1$, R, *a* and Γ are obtained. The fitted curves for different temperatures show the good agreement of the model with the experimental data in Fig. 4B. The square root of the fitting parameter $C_1$ and the mass density of D$_2$O have a linear relationship according to the expression of $C_1$ above. Extracting $C_1$ from the analysis, we obtain the absolute value of the density of D$_2$O in the pores by assuming it has the same density as bulk D$_2$O at 284±5 K. The fitted diameter of 16 Å of the water cylinder agrees closely with the diameter determined by nitrogen adsorption (15±1 Å). Table 1 and Fig. 1 show the D$_2$O density versus temperature. The plot shows a smooth transition of D$_2$O density from a higher value to a lower value. The higher value corresponds to the known density maximum of $D_2O$ at about 284 K. Most significantly, we find a density minimum situated at $T_{min}$=210±5 K with a value of 1.041±0.003 g/cm$^3$.

To confirm that the density minimum is a real physical phenomenon which does not depend on the fitting model we use, we plot the raw data together with the fitted curves in Fig. 4B. It can be directly seen from the graph that the intensity of the peak at 180 K is visibly higher than the intensity at 210 K. This demonstrates that the density minimum is not an artifact of the model fitting.

**Acknowledgement**

The research of Chen's group at MIT is supported by US DOE grant DE-FG02-90ER45429; of Mou's group by Taiwan NSC grant NSC95-2120-M-002-009; and of P. H. Poole by NSERC and the CRC Program. We appreciate the assistance of Dr. Y. Liu in the process of carrying out this experiment at NCNR. We would like to thank many fruitful conversations with Prof. H. Eugene Stanley on the subject of anomalies of water and the L-L critical point. We benefited from affiliation with EU Marie-Curie Research and Training Network on Arrested Matter.

**References**

1. Debenedetti, P. G. & Stanley, H.E., (2003) *Phys. Today* **56**, No. 6, 40.

2. Angell, C. A. & Kanno , H., (1976) *Science* **193**, 1121.

3. Rottger, K., Endriss, A. & Ihringer, J., *et al.*, (1994) *Acta Cryst.* **B50**, 644.

4. Angell, C.A., (2004) *Annu. Rev. Phys. Chem.* **55**, 559.

5. Angell, C.A., Bressel, R.D. & Hemmati, M., *et al.*, (2000) *Phys. Chem. Chem. Phys.* **2**, 1559.

6. Brovchenko, I., Geiger, A. & Oleinikova, A., (2001) *Phys. Chem. Chem. Phys.* **3**, 1567; (2003) *J. Chem. Phys.* 118, 9473; (2005) *J. Chem. Phys.* **123**, 044515.

7. Paschek, D., (2005) *Phys. Rev. Lett.* **94**, 217802.

8. Poole, P.H., Saika-Voivoid & Sciortino, I., F., (2005) *J. Phys: Condens. Matter* **17**, L431.

9. Mahoney, M. W. & Jorgensen, W. L., *J. Chem. Phys.* **112**, 8910 (2000)

10. Stillinger, F. H. & Rahman, A. , *J. Chem. Phys.* **60**, 1545 (1974)

11. Rusko, J. & Thurn, H., (1976) *J. Non-Cryst. Solids* **22**, 277.

12. Poole, P.H., Sciortino, F. & Essmann, U., *et al*., (1992) *Nature* **360**, 324.

13. Sen, S., Andrus, R.L. & Baker, D.E., *et al*., (2004) *Phys. Rev. Lett.* **93**, 125902.

14. Liu, L., Chen, S.-H. & Faraone, A. *et al*., (2005) *Phys. Rev. Lett.* **95**, 117802; Xu, L., Kumar, & P., Buldyrev, S. V., *et al*., (2005) *Proc. Natl. Acad. Sci. USA* **100**, 16558–16562

15. Kell, G. S. , (1967) *J. Chem. and Eng*. **12**, 66

16. Sastry, S., Debenedetti, P.G. & Sciortino, F., *et al*., (1996) *Phys. Rev. E* **53**, 6144.

17. Anisimov, M. A., Sengers, J. V. & Levelt Sengers, J. M. H., 'Near-critical Bahavior of Aqueous System' in Chapter 2, (2004, Elsevier) *Physical Chemistry in Water, Steam and Hydrothermal Solutions*, Eds., Palmer, D. A., Fernandez-Prini, R. & Harvey, A. H.

18. Private communication from H. Eugene Stanley.

19. Chen, S.-H., Lin, T. L. & Wu, C.F., in *Physics of Amphiphilic Layers*, Meunier, J., Langevin, D. & Boccara, N., Eds., (Springer-Verlag, 1987), pp. 241-252.

**Table 1. Fitted Model Parameters and the Measured Density of $D_2O$ as a Function of Temperature**

| T (K) | $C_1$ | R (Å) | $2\pi/a$ (Å$^{-1}$) | $\Gamma$ (Å$^{-1}$) | $\rho$ (g/cm$^3$) |
|---|---|---|---|---|---|
| 160 | 0.004269 | 8.09 | 0.287 | 0.0328 | 1.061 |
| 170 | 0.004237 | 8.09 | 0.287 | 0.0328 | 1.059 |
| 180 | 0.004201 | 8.09 | 0.287 | 0.0328 | 1.057 |
| 190 | 0.004111 | 8.09 | 0.287 | 0.0328 | 1.053 |
| 200 | 0.003984 | 8.09 | 0.287 | 0.0325 | 1.046 |
| 210 | 0.003889 | 8.09 | 0.287 | 0.0324 | 1.041 |
| 220 | 0.003907 | 8.09 | 0.287 | 0.0326 | 1.042 |
| 230 | 0.004006 | 8.09 | 0.287 | 0.0330 | 1.047 |
| 240 | 0.004244 | 8.09 | 0.288 | 0.0337 | 1.060 |
| 250 | 0.004483 | 8.09 | 0.288 | 0.0341 | 1.071 |
| 260 | 0.004819 | 8.09 | 0.288 | 0.0346 | 1.088 |
| 270 | 0.005017 | 8.09 | 0.288 | 0.0348 | 1.097 |
| 280 | 0.005168 | 8.09 | 0.288 | 0.0347 | 1.104 |
| 290 | 0.005203 | 8.09 | 0.288 | 0.0344 | 1.106 |

**Figure 1**

Average $D_2O$ density inside the 15±1 Å pore measured by SANS method as a function of temperature. A smooth transition of $D_2O$ density from the maximum value at 284±5 K to the minimum value at 210±5 K is clearly shown in the figure. The filled squares are the density data for bulk $D_2O$ taken from CRC handbook (1992).

**Figure 2**

Comparison of density versus temperature curves at ambient pressure for bulk liquid $D_2O$ (triangles) [CRC handbook], confined liquid $D_2O$ (solid circles) from this work, $D_2O$ ice Ih (solid squares) *(2),* and MD simulations of liquid TIP5P-E water (open diamonds) *(7).* The density values for the TIP5P-E model (which is parametrized as model of $H_2O$) have been multiplied by 1.1 to facilitate comparison with the behavior of $D_2O$.

**Figure 3**

SANS intensity distribution, as a function of Q, of MCM-41-S-15 plotted in a log-log scale. The blue solid line represents the interfacial scattering coming from the surfaces of the grains of the silica crystallites. Each scattering curve displays a power law region and a first order diffraction peak situated at $Q_1=0.287$ Å$^{-1}$. The diffraction peak is due to Bragg diffraction from the (01) plane of the 2-D hexagonal lattice made of parallel water cylinders contained in grains of the silica crystallites, which are oriented with the direction of the cylindrical axes parallel to the direction of the incident neutron beam. The height of the diffraction peak is proportional to the square of the difference of sld between $D_2O$ inside the cylindrical pore and the silica matrix. The height of the peak decreases steeply when the sample is cooled down in the temperature range from 280 K to 220 K indicating a steep decrease of density of $D_2O$. However the height increases significantly again in the temperature range from 210 K to 160 K where the density minimum was found by a detail analysis.

**Figure 4**

Model analysis of SANS intensity distribution.

(A) The blue circles show the SANS data (at 250K) with the contribution of the interfacial (surfaces of the grains) scattering subtracted. The red solid line represents the fitted curve using the model given in the text. The black line represents the structure factor $S(Q)$ of the 2-D hexagonal lattice. The green line represents the form factor $P(Q)$ of the cylindrical tube of $D_2O$ column. (To make the figure clearer, the magnitude of $S(Q)$ is multiplied by a factor $3 \times 10^{-4}$) .

(B) SANS data and their fitted curves (solid lines) for different temperatures. Four curves are selected to show the good agreement between the model analyses and the experimental data.

**Figure 1**

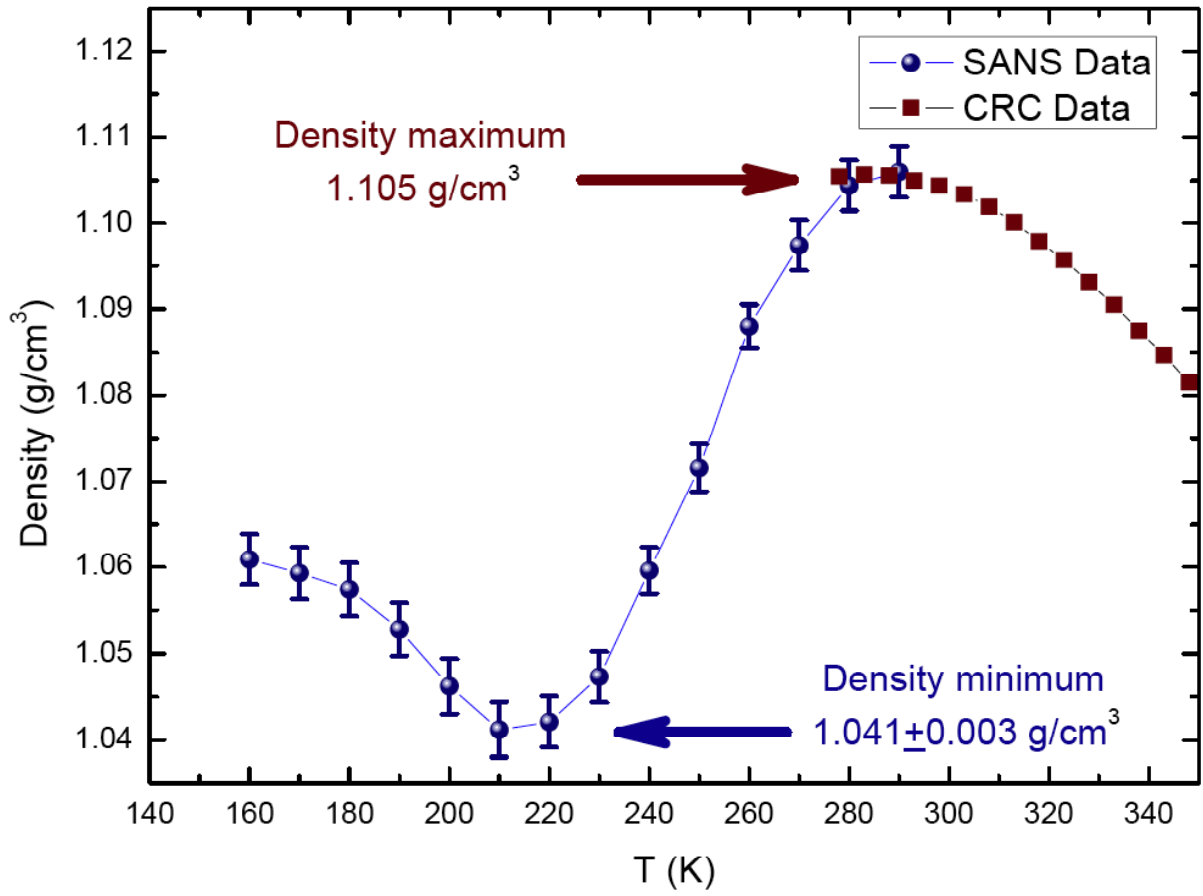

**Figure 2**

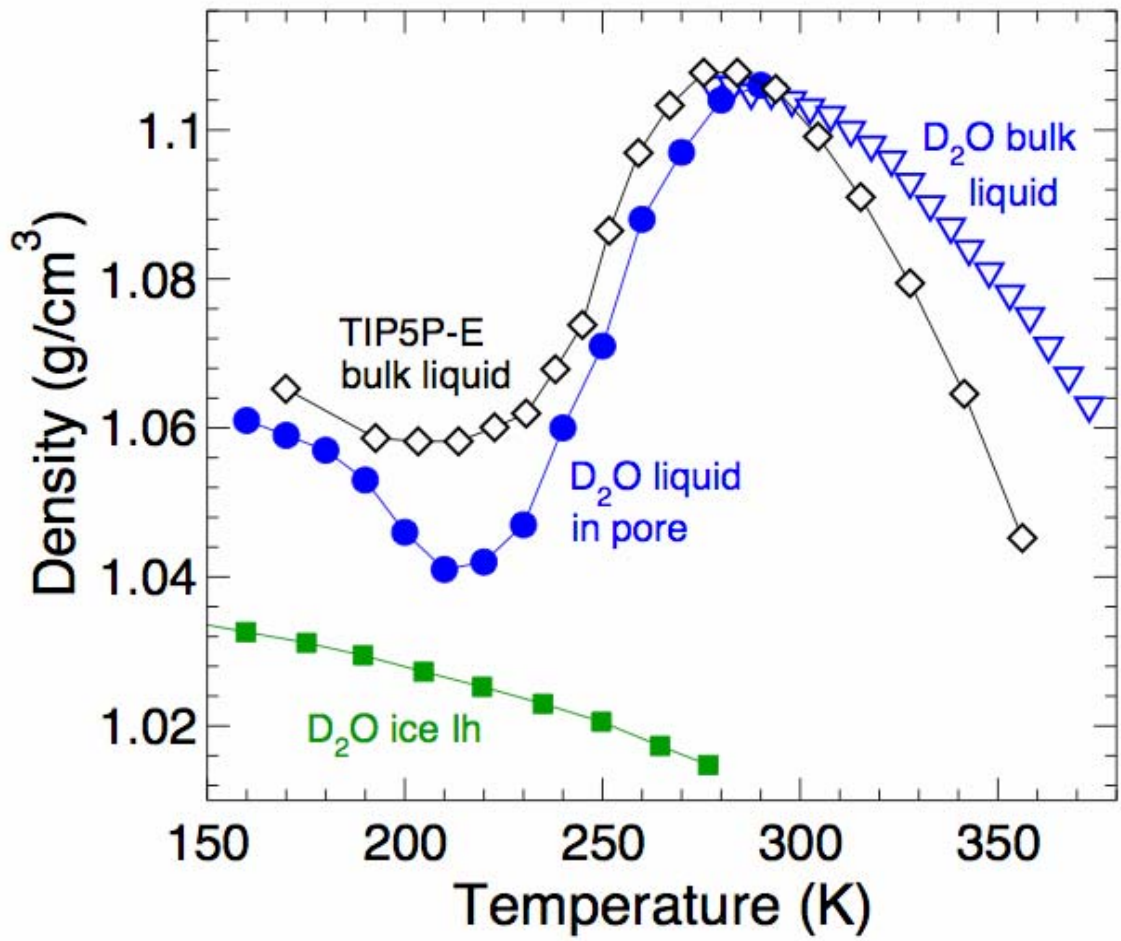

**Figure 3**

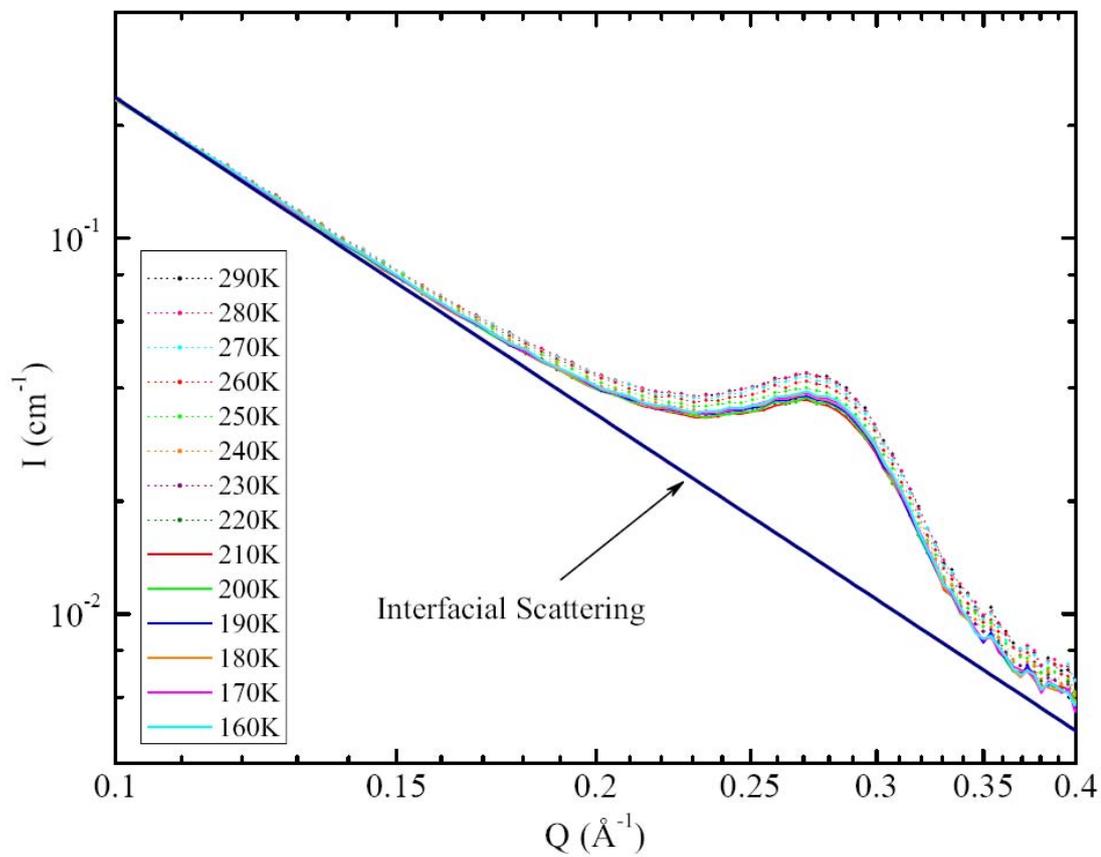

**Figure 4**

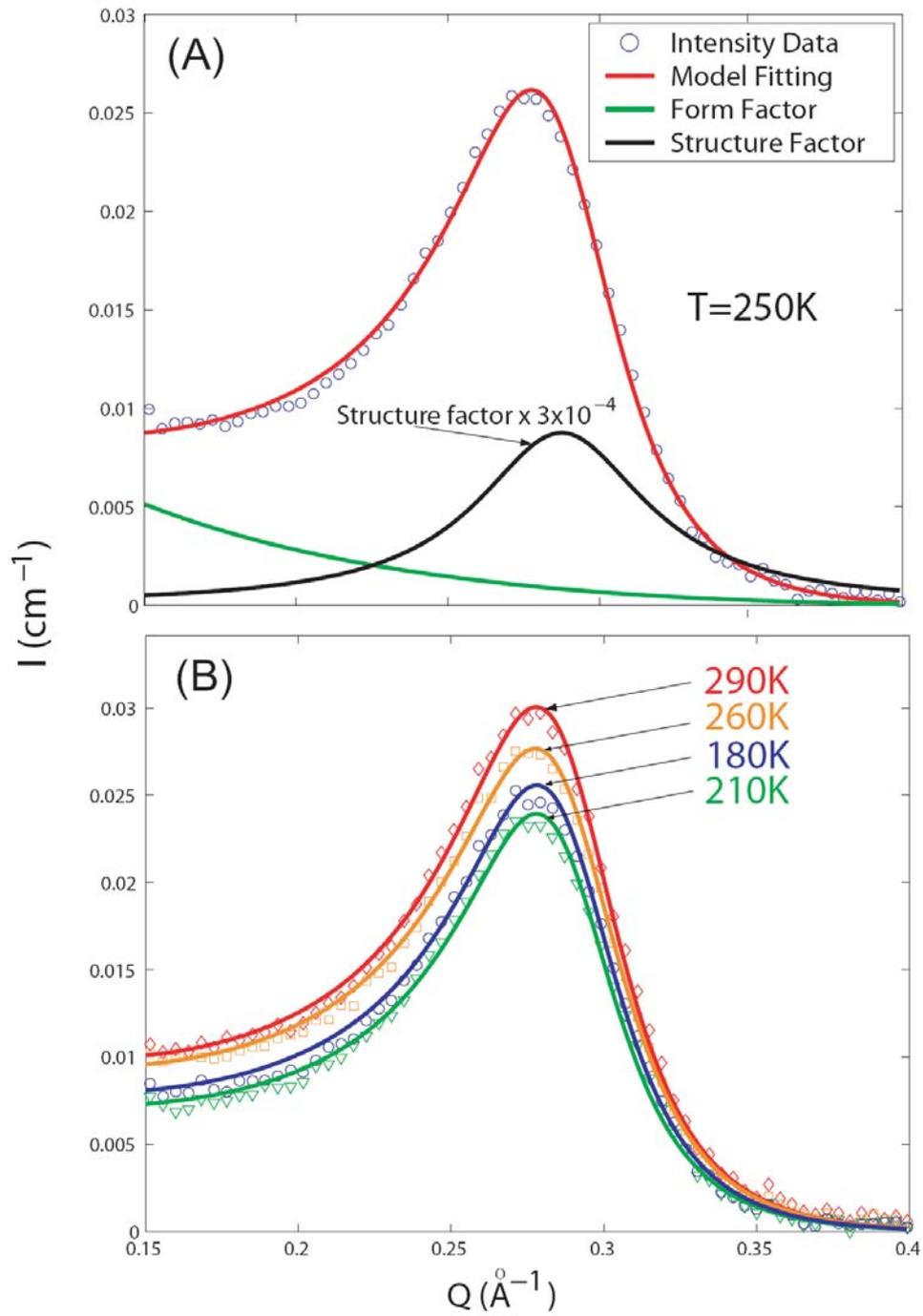